\documentclass[12pt]{article}
\usepackage{latexsym}
\usepackage{citesort}
\input epsf

\newcommand{\sect}[1]{\setcounter{equation}{0}\section{#1}}

\begin{document}

\title{Proliferation of de Sitter Space}

\author{{\sc Raphael Bousso}\thanks{\it
       bousso@stanford.edu}
       \\[1 ex] {\it Department of Physics}
       \\ {\it Stanford University}
       \\ {\it Stanford, California 94305-4060}
       }

\date{SU-ITP-98-24~~~~~hep-th/9805081}

\maketitle

\begin{abstract}
  
  I show that de~Sitter space disintegrates into an infinite number of
  copies of itself. This occurs iteratively through a quantum process
  involving two types of topology change. First a handle is created
  semiclassically, on which multiple black hole horizons form.  Then
  the black holes evaporate and disappear, splitting the spatial
  hypersurfaces into large parts.  Applied to cosmology, this process
  leads to the production of a large or infinite number of universes
  in most models of inflation and yields a new picture of global
  structure.
  
\end{abstract}

\pagebreak

\sect{Introduction}

\subsection{Semi-classical Perdurance of de Sitter Space}

de~Sitter space is the maximally symmetric solution of the vacuum
Einstein equations with a cosmological constant $\Lambda$.  It may be
visualized as a (3,1)-hyperboloid embedded in (4,1) Minkowski space.
In this spacetime, geodesic observers find themselves immersed in a
bath of thermal radiation~\cite{GibHaw77a} of temperature
$T=\sqrt{\Lambda/3}/(2\pi)$. This raises the question of stability,
which was investigated in 1983 by Ginsparg and Perry~\cite{GinPer83}.
They showed that de~Sitter space does not possess the classical Jeans
instability found in hot flat space. It does, however, possess a
semiclassical instability to a spontaneous topology change
corresponding to the nucleation of a black hole.\footnote{The process
  really describes a {\em pair} of black holes, in the sense that
  there will be two separate horizons. There will, however, be only
  one black hole interior. Later in this paper, situations with
  multiple black hole interiors will arise. Therefore any individual
  black hole interior, bounded by a pair of horizons, will be referred
  to here as a single black hole.} Geometrically this process
corresponds to the creation of a handle; it occurs at a rate of
$e^{-\pi/\Lambda}$ per horizon four-volume of size $9/\Lambda^2$.

When the black hole appears, it will typically be degenerate; that is,
it will have the same size as the cosmological horizon, and will be in
thermal equilibrium with it.  Ginsparg and Perry argued on an
intuitive basis that this equilibrium would be unstable, with quantum
fluctuations causing the black hole to be slightly smaller, and
hotter, than the cosmological horizon.  Then, presumably, the black
hole would start to evaporate and eventually disappear. They showed
that the time scale between black hole nucleations is vastly larger
than the time needed for evaporation (see also Ref.~\cite{BouHaw96}).
In this sense, de~Sitter space would ``perdure''.

In a collaboration with S.~Hawking, this argument was recently
confirmed~\cite{BouHaw97b}.  We used a model that includes the quantum
radiation in the s-wave and large $N$ limit, at one loop. We found, to
our surprise, that nearly degenerate Schwarzschild-de~Sitter black
holes anti-evaporate. But there is a different way of perturbing the
degenerate solution which leads to evaporation, and we found that this
mode would always be excited when black holes nucleate spontaneously.

\begin{figure}[htb]
  \hspace{.27\textwidth} \vbox{\epsfxsize=.46\textwidth
  \epsfbox{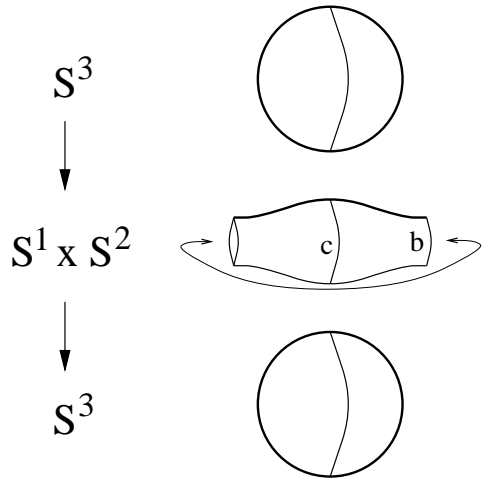}}
\caption%
{\small\sl Evolution of spacelike hypersurfaces of de~Sitter space
  during the creation and subsequent evaporation of a single black
  hole. The spontaneous creation of a handle changes the spatial
  topology from spherical ($S^3$) to toroidal ($S^1 \times S^2$) with
  constant two-sphere radius. (The double-headed arrow indicates that
  opposite ends of the middle picture should be identified, closing
  the $S^1$.) If the quantum fluctuations are dominated by the lowest
  Fourier mode on the $S^1$, there will be one minimum and one maximum
  two-sphere radius, corresponding to a black hole (b) and a
  cosmological horizon (c). This resembles a `wobbly doughnut' with
  cross-sections of varying thickness. As the black hole evaporates,
  the thinnest cross-section decreases in size. Finally the black hole
  disappears, i.e.\ the doughnut is pinched at its thinnest place and
  reverts to the original spherical topology.}
\label{fig-dissoc-1}
\end{figure}

\begin{figure}[htb]
  \hspace{.1\textwidth} \vbox{\epsfxsize=.8\textwidth
  \epsfbox{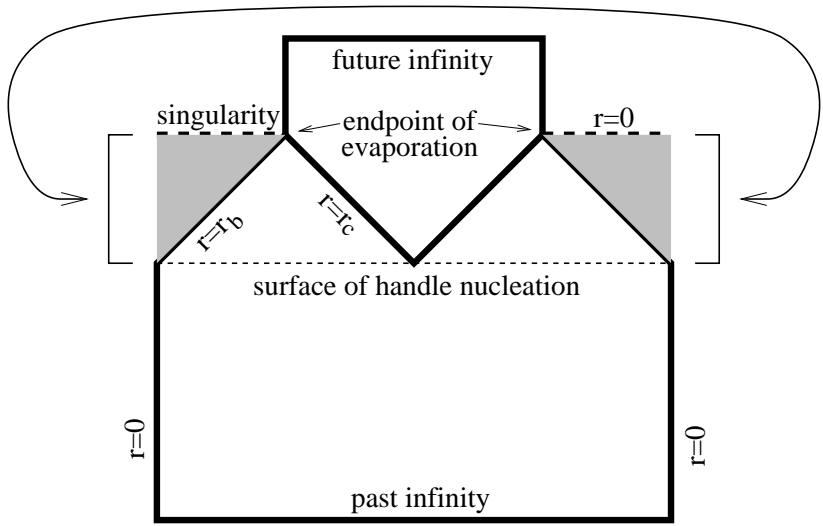}}
\caption%
{\small\sl Penrose diagram for the process depicted in
  Fig.~\ref{fig-dissoc-1}. The shaded region is the black hole
  interior. In the region marked by the square brackets the spatial
  topology is $S^1 \times S^2$, and opposite ends should be identified.
  The middle picture in Fig.~\ref{fig-dissoc-1} corresponds to the
  handle nucleation surface shown here. After the black hole
  evaporates, a single de~Sitter universe remains.}
\label{fig-penrose-1}
\end{figure}

The process of black hole creation and subsequent evaporation is shown
in Fig.~\ref{fig-dissoc-1}
(evolution of the spacelike sections) and
Fig.~\ref{fig-penrose-1} 
(causal structure). Crucially, for the case
of a single black hole, the topology reverts to the original de~Sitter
space after the process is completed.

\subsection{Proliferation}

In this paper, I consider different perturbations of the degenerate
black hole solution, which correspond to higher quantum fluctuation
modes. Occasionally, one such mode will dominate over lower modes,
leading to the presence of {\em multiple} pairs of apparent black hole
horizons in a single nucleation event, and eventually to the formation
of several black hole interiors. I will show that they evaporate, and
that space will disconnect at the event when a black hole finally
disappears. If more than one black hole is present, this leads to the
disintegration of the universe.

\begin{figure}[htb]
  \hspace{.2\textwidth} \vbox{\epsfxsize=.6\textwidth
  \epsfbox{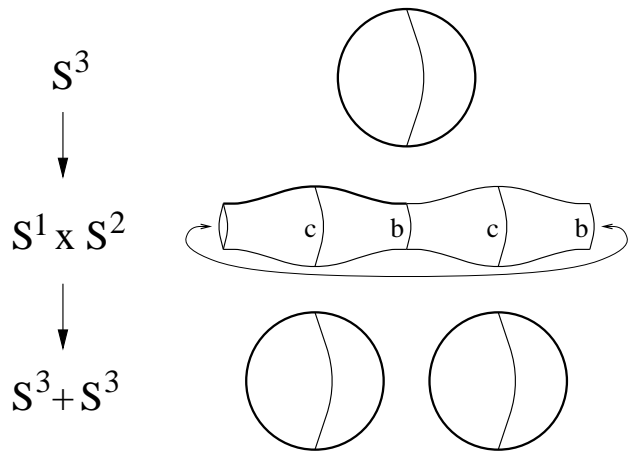}}
\caption%
{\small\sl Evolution of spacelike hypersurfaces of de~Sitter space
  during the creation of a handle yielding multiple black holes
  ($n=2$) and their subsequent evaporation. This should be compared to
  Fig.~\ref{fig-dissoc-1}. If the quantum fluctuations on the $S^1
  \times S^2$ handle are dominated by the second Fourier mode on the
  $S^1$, there will be two minima and two maxima of the two-sphere
  radius, seeding to two black hole interiors (b) and two
  asymptotically de~Sitter regions (c).  This resembles a `wobbly
  doughnut' on which the thickness of the cross-sections oscillates
  twice. As the black holes evaporate, the minimal cross-sections
  decrease.  When the black holes disappear, the doughnut is pinched
  at two places, yielding two disjoint spaces of spherical topology,
  the daughter universes.}
\label{fig-dissoc-n}
\end{figure}

\begin{figure}[htb]
  \hspace{.05\textwidth} \vbox{\epsfxsize=.9\textwidth
  \epsfbox{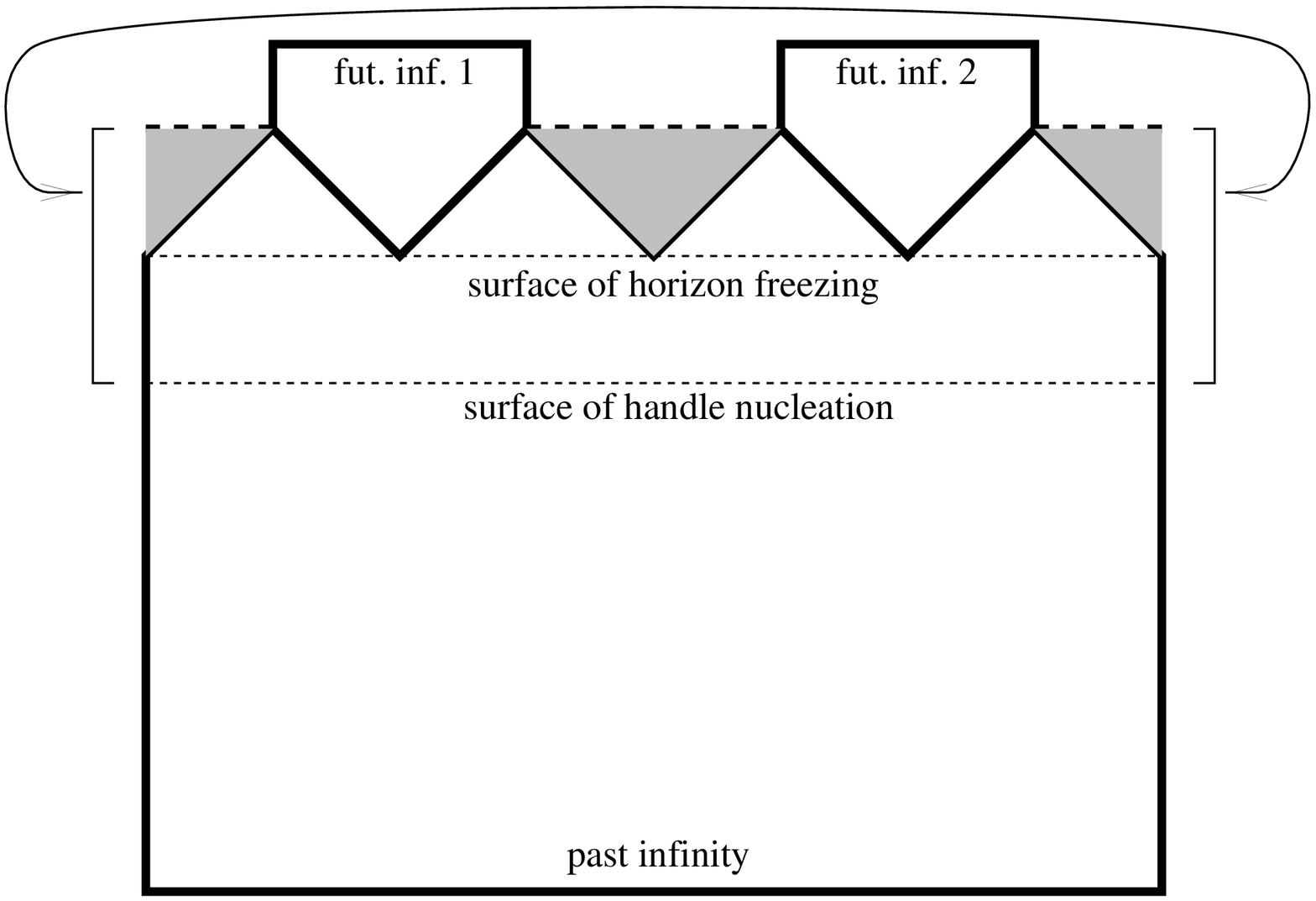}}
\caption%
{\small\sl Penrose diagram for the process depicted in
  Fig.~\ref{fig-dissoc-n}. The shaded regions are the two black hole
  interiors. In the region marked by the square brackets the spatial
  topology is $S^1 \times S^2$, and opposite ends should be identified.
  The middle picture in Fig.~\ref{fig-dissoc-n} corresponds to the
  horizon freezing surface shown here. After the black holes
  evaporate, two separate de~Sitter universes remain.}
\label{fig-penrose-n}
\end{figure}

This process is depicted in Fig.~\ref{fig-dissoc-n}, 
where the evolution of the space-like sections in a process of
multiple black hole creation and subsequent evaporation is shown. The
corresponding causal diagram is given in Fig.~\ref{fig-penrose-n}.
These pictures should be compared to the similar, yet much less
dramatic evolution in the case where only one black hole is created,
Figs.~\ref{fig-dissoc-1} and \ref{fig-penrose-1}.

The decay described here can be viewed as a sequence of topology
changes. The first topology change is the semiclassical creation of a
handle with multiple black hole horizons. It corresponds to a local
non-perturbative quantum fluctuation of the metric on the scale of a
single horizon volume. It will therefore happen independently in
widely separated horizon volumes of de~Sitter space. While the black
holes evaporate, the asymptotically de~Sitter regions between the
black holes grow exponentially. Thus, the second topology change,
corresponding to the final disappearance of the black holes, yields a
number of separate de~Sitter spaces which already contain
exponentially large spacelike sections. Because it occurs at such
large scales, the effect differs from many other considerations of
baby universes (see, e.g., Ref.~\cite{EQG}). Inside the daughter
universes, the handle-creation process will again occur locally. Thus
a single de~Sitter universe decomposes iteratively into an infinite
number of disjoint copies of itself, proliferating, as it were, by
budding.\footnote{If one starts with a single de~Sitter universe, the
  geometry resulting from the iterative disconnections will of course
  be connected if viewed as a four-dimensional manifold: at
  sufficiently early times, any two observers would have been in the
  same universe.  Only after the handles are nucleated, and the black
  holes evaporate, do the spatial sections fall apart. Spacetime
  decomposition can be defined in the following coordinate-independent
  way: For a geodesically complete Lorentzian four-manifold $\cal{M}$,
  consider the set $\cal{S}$ of all spacelike sections $S$ which have
  no boundaries (other than spatial infinity in the case of
  non-compact spaces). If there exists a pair $S_1, S_2 \in \cal S$
  such that $J^+(S_1) \cap J^+(S_2) = \emptyset$, the spacetime
  decomposes, and $S_1$ and $S_2$ are sections of different universes.
  Here $J^+(U)$ denotes the causal future of a set $U$.}

\subsection{Outline}

The geometry of black holes in de~Sitter space, and their spontaneous
nucleation, will be reviewed in Sec.~\ref{sec-cosmo}; this requires a
brief derivation of the Nariai solution, which corresponds to a
degenerate unstable handle attached to de~Sitter space.

In Sec.~\ref{sec-model}, a model will be introduced that includes
quantum radiation at one loop. It was used to investigate the behavior
of different perturbations of the Nariai solution, in
Ref.~\cite{BouHaw97b}.  There we considered only perturbations of the
two-sphere radius that corresponded to the lowest Fourier mode. But
the quantum model is general enough to deal also with higher mode
perturbations, for which the two-sphere size has $n$ maxima and $n$
minima around the $S^1$, with $n>1$.

In Sec.~\ref{sec-solutions}, I will present solutions for the
amplitude of such perturbations which are regular on the Euclidean
sector of the nucleation geometry. I will show that the condition of
regularity leads to the following Lorentzian evolution: The
perturbation oscillates while its wavelength is within a horizon
volume. When it leaves the horizon, it freezes, leading to the
formation of $n$ black holes, which are shown to evaporate. When a
black hole finally disappears, the spacelike section disconnects at
the evaporation point. The $S^1 \times S^2$ topology thus dissociates
into $n$ disjoint three-spheres.

Finally, in Sec.~\ref{sec-implications}, I discuss the cosmological
implications of this instability. During the inflationary era, it
causes the production of large new universes. In models of eternal
inflation, there will be an infinite number of separate universes.

\sect{Schwarzschild-de~Sitter Black Holes} \label{sec-cosmo}

The neutral, static, spherically symmetric solutions of the vacuum
Einstein equations with a cosmological constant $\Lambda$ are given by
the Schwarzschild-de~Sitter metric
\begin{equation}
ds^2 = - V(r) dt^2 + V(r)^{-1} dr^2 + r^2 d\Omega^2,
\label{eq-lsds}
\end{equation}
where
\begin{equation}
V(r) = 1 - \frac{2\mu}{r} - \frac{\Lambda}{3} r^2;
\end{equation}
$d\Omega^2$ is the metric on a unit two-sphere and $\mu$ is a mass
parameter. For $ 0 < \mu < \frac{1}{3} \Lambda^{-1/2} $, $ V $ has two
positive roots $r_{\rm c}$ and $r_{\rm b}$, corresponding to the
cosmological and the black hole horizons, respectively. For $\mu = 0$
there will be no black hole horizon and one obtains the de~Sitter
solution.

In the limit $\mu \rightarrow \frac{1}{3} \Lambda^{-1/2}$ the size of
the black hole horizon approaches the size of the cosmological
horizon, and the above coordinates become inadequate, since $V(r)
\rightarrow 0$ between the horizons. One may define the parameter
$\epsilon$ by
\begin{equation}
9 \mu^2 \Lambda = 1 - 3 \epsilon^2, \;\;
0 \leq \epsilon \ll 1. 
\end{equation}
Then the degenerate case corresponds to $ \epsilon \rightarrow 0 $.
New time and radial coordinates, $ \psi $ and $ \chi $, will be given
by
\begin{equation}
\tau = \frac{1}{\epsilon\sqrt{\Lambda}} \psi; \;\;\;
r = \frac{1}{\sqrt{\Lambda}}
\left[1 - \epsilon\cos\chi - \frac{1}{6} \epsilon^2 \right].
\label{eq-transformations}
\end{equation}
The black hole horizon lies at $ \chi = 0 $ and the cosmological
horizon at $ \chi = \pi $.  The new metric obtained from the
transformations is, to first order in $\epsilon$,
\begin{eqnarray}
ds^2 & = & - \frac{1}{\Lambda} \left( 1 +
                  \frac{2}{3}\epsilon\cos\chi
 \right) \sin^2\!\chi \; d\psi^2
     + \frac{1}{\Lambda} \left( 1 -
 \frac{2}{3}\epsilon\cos\chi \right)
                                 d\chi^2
\\ \nonumber
     & + & \frac{1}{\Lambda} \left( 1 -
 2\epsilon\cos\chi \right) d\Omega_2^2.
\label{eq-metric-eps}
\end{eqnarray}
It describes Schwarzschild-de~Sitter solutions of nearly maximal black
hole size.

In these coordinates the topology of the spacelike sections of
Schwarzschild-de~Sitter becomes manifest: $S^1 \times S^2$. (This is
why one can speak of handle-creation.) In these solutions, the radius
of the two-spheres, $r$, varies along the $S^1$ coordinate, $\chi$,
with the minimal (maximal) two-sphere corresponding to the black hole
(cosmological) horizon. In the degenerate case, for $ \epsilon =0 $,
the two-spheres all have the same radius.  The degenerate metric is is
called the ``Nariai'' solution; it is simply the direct product of
(1+1)-dimensional de~Sitter space and a two-sphere of constant radius.

The Euclidean de~Sitter solution can be obtained from the Lorentzian
metric, Eq.~(\ref{eq-lsds}), with $\mu=0$, by Wick-rotating the time
variable: $\tau = it$. This yields a Euclidean four-sphere of radius
$(\Lambda/3)^{-1/2}$. Its action is $-3\pi/\Lambda$. A regular
Euclidean sector also exists for the Nariai solution: With $ \epsilon
=0$ and $\xi = i \psi$, Eq.~(\ref{eq-metric-eps}) describes the
Euclidean metric corresponding to the direct product of two round
two-spheres of radius $\Lambda^{-1/2}$. This solution has a Euclidean
action of $-2\pi/\Lambda$.

To obtain the creation rate of the degenerate Schwarzschild-de~Sitter
solution on a de~Sitter background, one has to subtract the background
action and exponentiate minus the
difference~\cite{Col77,GinPer83,BouCha97}.  This yields the rate
$e^{-\pi/\Lambda}$. Clearly, the topological transition is strongly
suppressed unless the cosmological constant is of order the Planck
value. But this does not mean the process can be neglected; after all,
de~Sitter space, and many inflationary scenarios, contain an
exponentially large or infinite number of Hubble volumes.

\sect{One-loop Model} \label{sec-model}

The four-dimensional Lorentzian Einstein-Hilbert action with a
cosmological constant is given by:
\begin{equation}
S = \frac{1}{16 \pi} \int d^4\!x\, (-g^{{\rm IV}})^{1/2} \left[
 R^{{\rm IV}} - 2 \Lambda - \frac{1}{2} \sum_{i=1}^{N}
 (\nabla^{{\rm IV}} f_i)^2 \right],
\end{equation}
where $R^{{\rm IV}}$ and $g^{{\rm IV}}$ are the four-dimensional Ricci
scalar and metric determinant. The scalar fields $f_i$ are included to
carry the quantum radiation.

I shall consider only spherically symmetric fields and quantum
fluctuations. Thus the metric may be written as
\begin{equation}
ds^2 = e^{2\rho} \left( -dt^2 + dx^2 \right) + e^{-2\phi} d\Omega^2,
\label{eq-ssans}
\end{equation}
where $x$ is the coordinate on the $S^1$, with period $2\pi$. The angular
coordinates can be integrated out, which reduces the action to
\begin{equation}
S = \frac{1}{16\pi} \int d^2\!x\, (-g)^{1/2} e^{-2\phi} \left[
 R + 2 (\nabla \phi)^2 + 2 e^{2\phi} - 2
 \Lambda - \sum_{i=1}^{N} (\nabla f_i)^2 \right],
\end{equation}
In order to take the one-loop quantum effects into account, one should
find the classical solutions to the action $S+W^*$, where $W^*$ is the
scale-dependent part of the one-loop effective action for dilaton
coupled scalars~\cite{EliNaf94,MukWip94,ChiSii97,BouHaw97a,NojOdi97,%
Ich97,KumLie97,Dow98}:
\begin{equation}
W^* = - \frac{1}{48\pi} \int d^2\!x (-g)^{1/2} \left[ \frac{1}{2}
  R \frac{1}{\Box} R - 6 (\nabla \phi)^2 \frac{1}{\Box} R
  + 6 \phi R \right].
\end{equation}
As in Ref.~\cite{BouHaw97b}, the $(\nabla \phi)^2$ term may be
neglected.

This action can be made local by introducing an independent scalar
field $Z$ which mimics the trace anomaly. With the classical solution
$f_i=0$ the $f$ fields can be integrated out. This leads to the action
\begin{eqnarray}
S\!\!\! &
 =\!\!\! & \frac{1}{16\pi} \int d^2\!x\, (-g)^{1/2} \left[ \left( 
e^{-2\phi} + \frac{ N }{3} (Z - 6  \phi) \right) R
\right. \nonumber \\
& & \left. \mbox{\hspace{5em}}
 - \frac{ N }{6} \left( \nabla Z \right)^2
+ 2 + 2 e^{-2\phi} \left( \nabla \phi \right)^2
- 2 e^{-2\phi} \Lambda \right]\!.
\end{eqnarray}
In the large $N$ limit, the contribution from the quantum fluctuations
of the scalars dominates over that from the metric fluctuations. In
order for quantum corrections to be small, one should take $ N \Lambda
\ll 1 $.

Differentiation with respect to $t$ ($x$) will be denoted by an
overdot (a prime). For any functions $f$ and $g$, define:
\begin{equation}
\partial f\,\partial g  \equiv - \dot{f} \dot{g} + f' g',\ \ \ \ 
\partial^2 g \equiv - \ddot{g} + g'',
\end{equation}
\begin{equation}
\delta f\,\delta g \equiv \dot{f} \dot{g} + f' g',\ \ \ \
\delta^2 g \equiv \ddot{g} + g''.
\end{equation}
Variation with respect to $\rho$, $\phi$ and $Z$ yields the
following equations of motion:
\begin{equation}
- \left( 1 + N e^{2\phi} \right)
 \partial^2 \phi + 2
(\partial \phi)^2 + \frac{N}{6} e^{2\phi} \partial^2 Z +
e^{2\rho+2\phi} \left( \Lambda e^{-2\phi} - 1 \right) = 0;
\label{eq-m-rho}
\end{equation}
\begin{equation}
\left( 1 + N e^{2\phi}
 \right) \partial^2 \rho -
\partial^2 \phi + (\partial \phi)^2 + \Lambda e^{2\rho} = 0; 
\label{eq-m-phi}
\end{equation}
\begin{equation}
\partial^2 Z - 2 \partial^2 \rho = 0.
\label{eq-m-Z}
\end{equation}
There are two equations of constraint:
\begin{equation}
\left( 1 + N e^{2\phi} \right)
 \left( \delta^2 \phi - 2 \delta\phi\,\delta\rho \right) -
(\delta\phi)^2
= \frac{N}{12} e^{2\phi} \left[ (\delta Z)^2 + 2 \delta^2 Z
  - 4 \delta Z \delta\rho \right];
\label{eq-c1}
\end{equation}
\begin{equation}
\left( 1 + N e^{2\phi} \right)
 \left( \dot{\phi}' - 
\dot{\rho} \phi' - \rho' \dot{\phi} \right) - \dot{\phi} \phi'
= \frac{N}{12} e^{2\phi} \left[ \dot{Z} Z' + 2
\dot{Z}' - 2 \left( \dot{\rho} Z' + \rho'
  \dot{Z} \right) \right].
\label{eq-c2}
\end{equation}
From Eq.~(\ref{eq-m-Z}), it follows that $Z = 2\rho + \eta$, where
$\eta$ satisfies $\partial^2 \eta = 0$.  It was shown in
Ref.~\cite{BouHaw97b} that the remaining freedom in $\eta$ can be used
to satisfy the constraint equations for any choice of $\rho$, $
\dot{\rho} $, $\phi$ and $\dot{\phi}$ on an initial spacelike section.

\sect{Multi-Black Hole Solutions} \label{sec-solutions}

\subsection{Perturbation Ansatz}

In Sec.~\ref{sec-cosmo} it was described how one can describe the
process of handle-formation in de~Sitter space using instantons. It
was found that typically one obtains a degenerate Nariai solution as a
result of this topology change. In the coordinates of
Eq.~(\ref{eq-ssans}), it is given by
\begin{equation}
e^{2\rho} = \frac{1}{\Lambda_1} \frac{1}{\cos^2\! t}, \ \ \ 
e^{2\phi} = \Lambda_2,
\label{eq-nariai}
\end{equation}
Because of the presence of the quantum radiation, one no longer has
exactly $\Lambda_1 = \Lambda_2 = \Lambda$, but, to first order in $N
\Lambda$,
\begin{equation}
\frac{1}{\Lambda_1} =
  \frac{1}{\Lambda} \left( 1 + N \Lambda \right), \ \ \
\Lambda_2 =  \Lambda \left( 1 - \frac{N \Lambda }{3} \right)
\end{equation}
(see Ref.~\cite{BouHaw97b} for more details).

Quantum fluctuations will perturb this solution, so that the
two-sphere radius, $e^{-\phi}$, will vary slightly along the
one-sphere coordinate, $ x$. Decomposition into Fourier modes on the
$S^1$ yields the perturbation ansatz
\begin{equation}
e^{2\phi} =  \Lambda_2 \left[ 1 + 2 \epsilon
 \sum_n \left( \sigma_n(t) \cos nx + \tilde{\sigma}_n(t) \sin nx
 \right) \right],
\label{eq-fullpert-phi}
\end{equation}
where $\epsilon$ is taken to be small.

One does not lose much generality by assuming that the amplitude of
one mode dominates over all others.\footnote{This just means that the
  horizons will be evenly, rather than irregularly, spaced on the
  $S^1$. The constant mode ($n=0$) will be ignored here, since it does
  not lead to the formation of any black hole horizons, but to a
  topologically non-trivial, locally de~Sitter spacetime.} This will
most likely be the first Fourier mode ($n=1$), in which case one
obtains the usual single black hole investigated in
Ref.~\cite{BouHaw97b}, and no spatial decomposition takes place when
it evaporates. But occasionally, a higher mode will have a dominating
amplitude. It is reasonable to estimate the likelihood of such an
event to be of order $e^{-n^2}$.  For example, the second mode will
dominate in a few percent of handle-nucleation events. In any case,
the suppression of this condition is negligible compared to the rarity
of nucleating a handle in the first place. With this assumption, and a
trivial shift in the $S^1$ coordinate $x$, the perturbation ansatz
simplifies to
\begin{equation}
e^{2\phi} =  \Lambda_2 \left[ 1 + 2 \epsilon
 \sigma_n(t) \cos nx \right],
\label{eq-pert-phi}
\end{equation}

Following Ref.~\cite{BouHaw97b}, $\sigma_n$ will be called the {\em
  metric perturbation}.  A similar perturbation could be introduced
for $e^{2\rho}$, but it does not enter the equation of motion for
$\sigma_n$ at first order in $ \epsilon $. This equation is obtained
from the equations of motion for $\phi$, $\rho$, and $Z$, yielding
\begin{equation}
\frac{\ddot{\sigma_n}}{\sigma_n} = \frac{c(c+1)}{\cos^2\! t} - n^2,
\label{eq-m-si}
\end{equation}
where $c = 1 + 2 N \Lambda /9$ to first order in $N\Lambda$.

\subsection{Horizon Tracing}

In order to describe the evolution of the black hole, one must know
where the horizon is located. The condition for a horizon is
$(\nabla \phi)^2 = 0$. Eq.~(\ref{eq-pert-phi}) yields
\begin{equation}
\frac{\partial\phi}{\partial t} = \epsilon\, \dot{\sigma_n}
 \cos nx,\ \ \
\frac{\partial\phi}{\partial x} = - \epsilon\, \sigma_n\, n \sin nx.
\end{equation}
Therefore, there will be $2n$ black hole horizons, and $2n$
cosmological horizons, located at
\begin{eqnarray}
 x_{{\rm b}}^{(k)}(t) & = &
 \frac{1}{n} \left( 2\pi k + \arctan \left|
 \frac{\dot{\sigma}_n}{n\sigma_n} \right| \right), \\
 x_{{\rm b}}^{(n+k)}(t) & = &
 \frac{1}{n} \left( - 2\pi k + \arctan \left|
 \frac{\dot{\sigma}_n}{n\sigma_n} \right| \right), \\
 x_{{\rm c}}^{(l)}(t) & = & x_{{\rm b}}^{(l)}(t) + \frac{\pi}{n},
\label{eq-chib}
\end{eqnarray}
where $k = 0 \ldots n-1$ and $l = 0 \ldots 2n-1$.

To first order in $ \epsilon $, the size of the black hole horizons,
$r_{{\rm b}}$, is given by
\begin{equation}
r_{{\rm b}}(t)^{-2} = e^{2\phi[t, x^{(l)}_{{\rm b}}(t)]} = 
\Lambda_2 \left[ 1 + 2 \epsilon \delta(t) \right],
\label{eq-rb}
\end{equation}
where the {\em horizon perturbation} is defined to be
\begin{equation}
\delta \equiv \sigma_n \cos n x^{(l)}_{\rm b} =
\sigma_n \left( 1+
 \frac{\dot{\sigma_n}^2}{n^2\sigma_n^2} \right)^{-1/2}.
\label{eq-delta}
\end{equation}

To obtain explicitly the evolution of the black hole horizon radius,
$r_{{\rm b}}(t)$, one must solve Eq.~(\ref{eq-m-si}) for $
\sigma_n(t)$, and use the result in Eq.~(\ref{eq-delta}) to evaluate
Eq.~(\ref{eq-rb}). If the horizon perturbation grows, the black hole
is shrinking: this corresponds to evaporation. It will be shown below
that evaporation is indeed realized for perturbation solutions
$\sigma_n$ satisfying the condition of regularity on the Euclidean
instanton (the ``no-boundary condition'').

\subsection{Regular Solutions} \label{sec-regsolutions}

The metric of the Euclidean Nariai solution can be obtained from
Eq.~(\ref{eq-nariai}) by taking $\tau=it$. After rescaling Euclidean
time as $\sin u = 1/\cosh \tau$, it takes the form
\begin{equation}
ds^2 =  \frac{1}{\Lambda_1} \left( du^2 +
  \sin^2\! u\, d x^2 \right) +  \frac{1}{\Lambda_2} d\Omega^2.
\end{equation}
This $S^2 \times S^2$ instanton describes the spontaneous nucleation
of a degenerate handle in de~Sitter space.  The nucleation path runs
from the South pole of the first two-sphere, at $u=0$, to $u=\pi/2$,
and then parallel to the imaginary time axis ($u=\pi/2+iv$) from $v=0$
to $v=\infty$. This can be visualized geometrically by cutting the
first two sphere in half, and joining to it a Lorentzian
$1+1$-dimensional de~Sitter hyperboloid.

In order to see the effect of quantum fluctuations, one must find
solutions to the equation of motion for the perturbation amplitude,
Eq.~(\ref{eq-m-si}). The solutions have to be regular everywhere on
the nucleation geometry. In particular, they must vanish on the south
pole. This condition selects a one-dimensional subspace of the
two-dimensional solution space of the second-order equation. For any
$n \geq 1$, the family of solutions, parametrized by a real prefactor
$A$, is given by
\begin{equation}
\sigma_n(u) = A e^{i(c-n) \pi/2} \left( \tan \frac{u}{2} \right)^n
  \left( n + \cos cu \right).
\end{equation}
The phase is chosen such that $\sigma_n$ will be real at late
Lorentzian times, when measurements can be made. The solution is exact
for $c=1$ (which corresponds to no quantum matter) and is an excellent
approximation for $N \Lambda \ll 1$, $n \geq 2$.

The Lorentzian behavior of this solution can be obtained by
substituting $u = \pi/2 + iv$.  While the $n^2$ term dominates on the
right hand side of Eq.~(\ref{eq-m-si}), one would expect the
perturbation to oscillate. Indeed, it is easy to show that the above
solution oscillates until $v \approx \mbox{arcsinh}\, n$, during which
time it undergoes approximately $(n \arctan n)/2\pi - 1/8$ cycles.

The Lorentzian evolution of the background metric is characterized by
the exponential growth of the $S^1$ radius, while both the $S^2$ size
and the horizon radius are practically constant. Perturbations with $n
\geq 2$ initially have a wavelength smaller than the horizon and
therefore represent not black holes, but mere ripples in the metric.
This explains the initial oscillatory behavior. But the wavelength
grows with the $S^1$ radius until the perturbations leave the horizon.
Then they freeze, seeding the gravitational collapse of the
two-spheres in the $n$ regions between the $2n$ black hole horizons,
and the exponential growth of the two-spheres in the $n$ regions
between the $2n$ cosmological horizons.

Due to the exchange of quantum radiation between the horizons, the
black holes evaporate and shrink, while the cosmological horizons
become larger.  This follows from the behavior of $\delta(v)$ at late
Lorentzian times, when $\sigma_n(v) = A \sinh cv$.  Indeed, from
Eq.~(\ref{eq-delta}) one finds that the horizon perturbation grows:
\begin{equation}
\delta(v) =
 \frac{An}{c} \exp \left( \frac{2 N \Lambda}{9} v \right).
\end{equation}
The behavior of the multiple black holes is thus similar to that of
the single black hole investigated in Ref.~\cite{BouHaw97b}. This
result was confirmed for conformal scalars in Ref.~\cite{NojOdi98a}.
Recently it was claimed that for different types of quantum matter,
some of the spontaneously created handles may be stable at least
initially~\cite{NojOdi98b}. For the arguments given here, however, it
is sufficient that a fraction of handles develop multiple evaporating
black holes.

The perturbative description breaks down when the black holes become
much smaller than the cosmological horizons. But then the cosmological
horizon will have a negligible influence on the black hole it
surrounds, which will behave like a neutral black hole immersed in
asymptotically flat space, evaporating at an ever-increasing rate.
While standard physics breaks down at the endpoint of the evaporation
of an uncharged black hole, it seems reasonable to assume that it will
disappear altogether in a final Planckian flash of radiation.

Each black hole on the $S^1$ is surrounded by two cosmological
horizons, beyond which lie intermediate regions bounded by
cosmological horizons surrounding the two neighboring black holes (see
Fig.~\ref{fig-penrose-n}).  As the regions between the black hole
horizons undergo gravitational collapse, these regions between pairs
of cosmological horizons undergo a similar run-away, corresponding to
the exponential expansion into asymptotically de~Sitter universes.
While the black holes evaporate, they grow extremely large. When a
black hole finally disappears, the two asymptotic de~Sitter regions on
either side of it will disconnect.  If there is only one black hole on
the entire $S^1$, these ``two'' asymptotic de~Sitter regions are
merely opposite ends of the same region.  Then the universe does not
actually decompose into two parts, but merely reverts to its original,
trivial $S^3$ topology.  If, however, there are $n \geq 2$ black holes
on the $S^1$, there will be $n$ distinct asymptotic de~Sitter regions.
As all the black holes evaporate, these large regions will pinch off,
and become $n$ separate de~Sitter universes. Each of the daughter
universes will themselves harbor handle-creation events, so the
disintegration process continues ad infinitum.

In this sense, de~Sitter space proliferates.

\sect{Cosmological Implications} \label{sec-implications}

In most models of inflation, the universe undergoes a period of
exponential expansion, driven by the vacuum energy $V(\phi)$ of a
scalar field. While the evolution is vacuum dominated, the universe
behaves like de~Sitter space with an effective cosmological constant
proportional to $V$. Typically, the field $\phi$ rolls down slowly,
corresponding to a slow decrease of the effective cosmological
constant, until it reaches the minimum of its potential. There $V=0$,
and inflation ends. Because of the slowness of this change, the
universe looks locally de~Sitter during inflation.

The handle-creation effect is a local event taking place on the scale
of a horizon volume; on this scale, the change in $\phi$ is usually
negligible. This is true for open~\cite{HawTur98} as well as closed
models.  Therefore handle creation takes place during inflation and
can be described by the same methods that were used above for the
case of a fixed cosmological constant~\cite{BouHaw95,BouHaw96}. Then
multiple black holes can form and the universe will disintegrate when
they evaporate.

It is easy to check that for most models of chaotic inflation, with
power-law inflaton potentials, the total number of handle-creation
events during inflation will be exponentially large. A non-negligible
fraction of these handles will evolve into multiple black holes and
eventually induce the disintegration of the inflating universe. Thus
the inflationary era starts in a single universe but ends in an
exponentially large number of disconnected universes, each of which
enters into a radiation dominated phase.

According to most inflationary models, the universe is vastly larger
than the present horizon. The proliferation effect I have discussed
renders our position even more humble: Not only do we live in an
exponentially small part of the universe, compared to its global size;
but our universe is only one of exponentially many disconnected
universes, all originating from the same small region in which
inflation started.

So far it has been assumed that the inflaton field rolls down slowly
according to its classical equations of motion.
Linde~\cite{Lin86a,LinLin94} has shown, however, that for sufficiently
large values of the inflaton field, its random quantum fluctuations
will dominate over the classical decrease. Every Hubble time, $e^3
\approx 20$ new horizon volumes are produced. If the quantum jumps of
$\phi$ dominate, the effective cosmological constant will increase in
about 10 of these new volumes, and decrease in the other 10. Thus
inflation continues forever in some regions, and ends only where the
inflaton field happens to jump down into the regime in which it must
decrease classically.

This idea, called ``eternal inflation'', has profound consequences for
the global structure of the universe. It implies that inflation never
ends globally and leads to a stationary overall stochastic state of
the universe in some models.  Crucially, in models allowing eternal
inflation, there will be an infinite number of handle-creation events,
and correspondingly, the universe will disintegrate into infinitely
many parts. Inflation continues forever, and the production of new
universes never ceases. The combination of eternal inflation with the
disintegration effect, then, has led to a picture of global structure
in which the number of separate universes is unbounded.

\sect{Conclusions}

de~Sitter space proliferates by disintegrating into separate, large
copies of itself. This occurs through a decay process involving the
formation and evaporation of multiple black holes. The process
consists of several steps:
\begin{itemize}
\item{The spontaneous creation of a handle, changing the spatial
    topology from $S^3$ to $S^1 \times S^2$. This process is
    suppressed by a factor of $e^{-\pi/\Lambda}$ and can be described
    semiclassically using gravitational instantons.}
\item{The breaking of the degeneracy of the handle geometry by random
    quantum fluctuations. In the saddlepoint solution, the radius of
    the two-spheres is independent of the $S^1$ coordinate.  If the
    $n$-th Fourier mode perturbation dominates, $n$ black hole
    interiors will form after the $S^1$ has expanded sufficiently.}
\item{The evaporation of the $n$ black holes. By including one-loop
    quantum radiation in the s-wave and large $N$ approximation, one
    finds that black holes nucleated semi-classically in de~Sitter
    space evaporate. Meanwhile, exponentially large de~Sitter regions
    develop beyond the cosmological horizons.}
\item{The disconnection of the spacelike hypersurfaces. When the black
    holes finally disappear, the toroidal $S^1 \times S^2$ topology is
    pinched at the evaporation endpoint: The $S^2$ radius becomes zero
    there and the hypersurface dissociates. After all the black holes
    evaporate, this leaves $n$ exponentially expanding three-spheres,
    corresponding to $n$ separate de~Sitter universes.}
\end{itemize}
The process repeats in the resulting fragments and continues
indefinitely, producing infinitely many distinct universes.

According to the inflationary paradigm, the most successful theory of
primordial cosmology, our universe went through a period of
de~Sitter-like expansion before the onset of the radiation and matter
dominated eras. Therefore the process described in this paper is
relevant to the global structure of the universe: It means that we
live in one of many universes that originated in the same primordial
inflationary region. There is a large class of models that lead to
eternal inflation; in this case, we inhabit one of an infinite number
of separate universes produced from a single region.

The process I have described lends itself to various generalizations,
further enriching our picture of the global structure of the universe.
One could consider, for example, the creation of handles carrying
magnetic flux~\cite{MelMos89,BouHaw96}. The resulting black holes
could not evaporate completely, resulting in a network of large
de~Sitter bubbles connected through thin extremal black hole throats.

\subsection*{Acknowledgments}

The notion of multiple black hole formation on a handle originated in
a conversation with Stephen Hawking. I would like to thank him and
Andrei Linde for many useful discussions. I am grateful to Andrew
Chamblin, Andrei Linde, and Malcolm Perry for comments on a draft of
this paper. This work was supported by NATO/DAAD.

\bibliographystyle{board}
\bibliography{all}

\end{document}